\documentclass[12pt,letterpaper]{article}
\usepackage{float}
\usepackage{multirow, dcolumn}
\usepackage{ucs,amssymb,amsmath,mathtools}
\usepackage[utf8]{inputenc}
\usepackage{fontenc}
\usepackage{nicefrac}
\usepackage{graphicx}
\usepackage{changepage}
\usepackage[usenames,pdftex,dvips]{color,xcolor}
\usepackage[pdftex,colorlinks]{hyperref,ragged2e}
\usepackage{apacite}
\usepackage{booktabs}
\usepackage{threeparttable}
\usepackage{pdflscape}
\usepackage[
singlelinecheck=false 
]{caption}
\usepackage{soul}

\usepackage[margin=1in]{geometry}
\usepackage{setspace}
\usepackage[normalem]{ulem}

\newcommand{\Nb}{ \nobreak}

\DeclareMathOperator{\OR}{OR}
\DeclareMathOperator{\RR}{RR}
\DeclareMathOperator{\Var}{Var}
\DeclareMathOperator{\LLC}{LLC}
\DeclareMathOperator{\sech}{sech}
\DeclareMathOperator*{\argmin}{argmin}
\DeclareMathOperator\arctanh{arctanh}
\def\Asym{\stackrel{H_0}{\sim}}

\title{A new measure for the analysis of epidemiological associations: Cannabis use disorder examples}
\date{}
\begin{document}
\maketitle
\thispagestyle{empty}
\begin{flushleft}
  \vskip -10ex
  \begin{center}
    \textbf{Short Title:} Inference using normalized odds ratio\\
    \end{center}
\vskip 5ex
  \author{Olga A. Vsevolozhskaya$^1$, Karl C. Alcover$^2$, James C. Anthony$^{3}$, Dmitri V. Zaykin$^{4*}$}\\
  $^1$Department of Biostatistics, University of Kentucky, Lexington, KY, USA\\
  $^2$Behavioral Health Innovations, Washington State University, Pullman, WA, USA\\
  $^3$Department of Epidemiology and Biostatistics, East Lansing, MI, USA\\
  $^4$National Institute of Environmental Health Sciences, National Institutes of Health, Durham, NC, USA
  \vskip 2ex
  \textbf{Correspondence}$^*$: Dmitri V. Zaykin, Biostatistics and Computational Biology Branch, National Institute of Environmental Health Sciences, National Institutes of Health, P.O. Box 12233, Research Triangle Park, NC 27709, USA. Email address: dmitri.zaykin@nih.gov
   \vskip 1ex
   \textbf{Co-Correspondence}$^\dagger$: James C. Anthony, Department of Epidemiology and Biostatistics,  College of Human Medicine,  Michigan State University,  909 Fee Road,  East Lansing, MI 48824-1030, USA. Email address: janthony@msu.edu
\vskip 1ex
 \textbf{Acknowledgments}: The authors would like to thank Professor Sander Greenland for helpful comments and suggestions, and Gabriel Ruiz for his contribution to establishing the LLC--log(OR) connection during his Summer of Discovery undergraduate internship at NIEHS.
\end{flushleft}

\clearpage
\setcounter{page}{1}
\section*{Abstract}
Analyses of population-based surveys are instrumental to research on prevention and treatment of mental and substance use disorders. Population-based data provides descriptive characteristics of multiple determinants of public health and are typically available to researchers as an annual data release. To provide trends in national estimates or to update the existing ones, a meta-analytical approach to year-by-year data is typically employed with ORs as effect sizes. However, if the estimated ORs exhibit different patterns over time, some normalization of ORs may be warranted. We propose a new normalized measure of effect size and derive an asymptotic distribution for the respective test statistic. The normalization constant is based on the maximum range of the standardized log(OR), for which we establish a connection to the Laplace Limit Constant. Furthermore, we propose to employ standardized log(OR) in a novel way to obtain accurate posterior inference. Through simulation studies, we show that our new statistic is more powerful than the traditional one for testing the hypothesis OR=1. We then applied it to the United States population estimates of co-occurrence of side effect problem-experiences (SEPE) among newly incident cannabis users, based on the the National Survey on Drug Use and Health (NSDUH), 2004-2014.

\vskip 2ex
\noindent \textbf{Keywords}: standardized coefficient; posterior interval estimation; cannabis; risk factors
\clearpage

\section{INTRODUCTION}
As of May 2020, more than 30 state-level jurisdictions of the United States (US) have liberalized cannabis control policies. Even so, federal law remains unchanged and is compliant with international cannabis prohibition treaties. Judged by this standard, cannabis qualifies as the most commonly used internationally regulated drug (IRD).

Social attitude surveys about perceived risk disclose low-level risk perceptions about cannabis, relative to risk perceptions about cocaine, heroin, and other IRD \cite{johnston2016monitoring}. The US Surgeon General thinks otherwise and recently released an advisory report intended to increase awareness of evidence about potential harms of cannabis, which include concerns about cannabis use by pregnant women, and by children and adolescents, with a specific focus on the developing adolescent brain \cite{volkow2017risks, volkow2014adverse}. The report also drew attention to addictive processes that may lead to a severe cannabis use disorder (CUD), especially when cannabis products contain high concentrations of delta-9-tetrahydrocannabinol (THC).

One way to address issues associated with adverse psychiatric and general medical outcomes of cannabis use is to focus attention on promising targets of early CUD screening, as well as interventions to `prevescalate' and disrupt newly incident CUD processes. If successful, focused attention can disclose facets of a CUD `target product profile' (TPP), i.e., a set of clearly defined features that indicate an early progression to CUD. Optimally, TPP can be developed backward from a potential therapeutic target toward an intermediate prodromal stage with $<100$\% predictive value of a positive test for the target disease of interest (e.g., CUD). Alternative TPP processes start with a molecule being worked up in medicinal chemistry or pharmaceutics laboratories, with evidence supportive of both adequate safety and some `side effect' indicative of therapeutic value. In our TPP example, we have no specific molecule, device, or intervention in mind. Instead, we offer an array of potential pre-CUD therapeutic targets manifest as syndromes in the sense of a statistically tangible co-occurrence of pairs of indicators. We posit that the co-occurring indicators might become promising potential therapeutic targets of potential utility in an accelerated CUD medication development process.

Ordinarily, epidemiologists might turn to the familiar odds ratio (OR) to investigate the patterns of co-occurrence. ORs have useful properties, such as invariance across sampling designs (e.g., case-control or prospective), and straightforward interpretation of estimates from logistic regression modeling. The log-transformed odds ratios, log(OR), also have analytically attractive features such as asymptotic distribution of log(OR) rapidly converging to a normal distribution with an increasing sample size. 

In this paper, we propose a new normalized measure of effect size, $-1<\gamma^{\prime}<1$, and derive an asymptotic distribution for the respective test statistic to test the null hypothesis OR=1. $\gamma^{\prime}$ is derived by considering the lower and the upper bounds on the possible values of the standardized log(OR), i.e., $\gamma=\log(\OR) /  \sigma$, and by showing that $\gamma$ can not exceed the Laplace Limit Constant (LLC). Then, using the LLC as a normalizing constant, we propose a new measure of the effect size, $\gamma' = \nicefrac{\gamma}{\LLC}$, that is varying between -1 and 1. We derive an asymptotic distribution of $\gamma'$ and show, via simulation experiments, that the new association statistic based on the $\gamma'$ distribution promises to be a more discriminating and statistically more powerful alternative to OR. We further show how the standardized $\log(\OR)$ can be utilized to accurately approximate posterior inference for the the newly proposed $\gamma^{\prime}$, and utilize this property to investigate year-by-year patterns of co-occurrence of side effect problem-experiences (SEPE) associated with accelerated CUD development process. 

\section{MATERIALS AND METHODS}
\subsection{Subjects and Measures}

For our epidemiological estimates from the US National Surveys on Drug Use and Health (NSDUH), the population under study was specified as the civilian residents of non-institutional dwelling units (DU) of the 50 United States and the District of Columbia, age 12 years and older. Each year from 2004-2014, the NSDUH field research staff used a multi-stage area probability sampling approach to draw a new sample of designated respondents (DR) within each DU. Each DR was recruited and asked to complete an audio computer-assisted self-interview (ACASI), organized as a fixed sequence of standardized survey items within a series of modules on drug use and health. One module covers month and year of first cannabis use, as well as cannabis side effect problems and experiences (SEPE). Methodological details on the survey have been published in many online reports and open access articles \cite{samhsa04, samhsa14, samhsa_codebook, alcover2019side}.

Each year's protocol for human subjects protection is reviewed by a cognizant institutional board. In its early years, NSDUH was able to secure participation levels above 70\%, but in recent years participation has dropped to the 55\%-65\% range. The NSDUH public-use data files provide a single analysis weight designed to take into account the selection probabilities as well as post-stratification adjustment factors intended to bring the analysis-weighted estimates into balance with the US Census marginal distributions. Our work is based on the NSDUH Public-use Data Analysis System (PDAS) downloadable datasets for which each year’s sample has included between 50,000 and 60,000 participants \cite{samhsa04, samhsa14, samhsa_codebook, alcover2019side}.

Designed and conducted as an annual cross-sectional survey with successive replication samples of the US population, the cannabis module's month-by-month data on the first occasion of cannabis use make it possible to focus on the set of newly incident cannabis users observed within the interval of about 1 to 90 days after first cannabis use. After aggregation across the years, this set encompassed $n = 3,710$ newly incident cannabis users, mainly age 12-to-29-years-old, for whom 17 cannabis side effect problems and experiences (SEPE) were assessed. As discussed elsewhere, the cannabis module was designed with an assumption of no SEPE unless cannabis was used six or more times within the interval since cannabis onset \cite{alcover2019side}. Table 1 lists the SEPEs in the order determined by incidence rate estimates \cite{alcover2019side}. The most frequently observed SEPE is `wanting or trying to cut down or stop using,' which affected more than 50\% of the newly incident cannabis users within $\sim90$ days after 1st use. In contrast, `continuing to use despite physical problems'' affected few ($\sim$1\%). The prior work showed estimated odds ratios for SEPE-SEPE pairs across a range from $1.0$ to $>10.0$ \cite{alcover2019side}. We have re-approached the analysis of these data and derived estimates for the standardized logarithm of the odds ratio for comparison with the original odds ratio estimates.

\subsection{Bounds for the standardized logarithm of odds ratio} \label{bounds}
We shall assume for now that epidemiological data is summarized by a 2$\times$2 table as:
\begin{table}[ht!]
\centering
\begin{tabular}{ccc}
\hline
&\multicolumn{2}{l}{Exposure status}\\
\cline{2-3} \rule{0pt}{3ex}
Disease status & $E$ & $\bar{E}$ \\
\hline \rule{0pt}{3ex}
$D$ & $n_{11}=n_D\hat{p}$ & $n_{12}=n_D(1-\hat{p})$\\
$\bar{D}$ & $n_{21}=n_{\bar{D}}\hat{q}$ & $n_{22}=n_{\bar{D}}(1-\hat{q})$\\
\hline
\end{tabular}
\label{tab1}
\end{table}

\noindent
where $n_{11}+n_{12} = n_D$ is the number of cases; $n_{21}+n_{22} = n_{\bar{D}}$ is the number of controls; and the number of exposed subjects is $n_{11}+n_{21}$. When sampling is random with respect to exposure, sample proportions $\hat{p}=\nicefrac{n_{11}}{n_D}$ and $\hat{q}=\nicefrac{n_{21}}{n_{\bar{D}}}$ are estimates of the population probabilities of exposure among cases and among controls, respectively, $p = \Pr(E|D)$ and $q = \Pr(E|\bar{D})$. Then, the effect of exposure on an outcome can be measured by the odds ratio, OR, defined as:
\begin{eqnarray*}
   \OR &=& \frac{ p/(1-p) }{ q/(1-q)  } \\
 &=& \frac{ \Pr\left(E \mid D\right) \left[1-\Pr\left(E \mid \bar{D}\right)\right]  }{\Pr\left(E \mid \bar{D}\right) \left[1-\Pr\left(E \mid D\right)\right]}.
\end{eqnarray*}
To study influence of various risk factors on the outcome, one can test the null hypothesis $H_0: \OR = 1$, or equivalently $H_0$: log(OR)=0. The logarithmic transformation is advantageous because of the bounded and asymmetric nature of OR (it can not take negative values) and also due to the fact that the distribution of log(OR) quickly converges to normality. Then, the classical test statistic is defined as: 
\begin{eqnarray*}
   Z & = & \frac{\log\left(\widehat{\OR}\right)} {\sqrt{\sum{  \nicefrac{1}{n_{ij}}  }}} \\ 
 & = & \sqrt{N} \, \, \frac{\log(\widehat{\OR})} {\hat{\sigma}}, \quad \text{where} \quad
  \hat{\sigma} = \sqrt{\frac{1}{\hat{w}} \frac{1}{\hat{p}(1-\hat{p})} + \frac{1}{1-\hat{w}} \frac{1}{\hat{q}(1-\hat{q})}} \,\,\, ,
\end{eqnarray*} 
and $\hat{w}$ is the sample proportion of cases, $\nicefrac{n_D}{N}$. 
The corresponding population parameter can be written as:
\begin{eqnarray}  
    \sigma^2 &=& \frac{1}{w} \frac{1}{p(1-p)} + \frac{1}{(1-w)} \frac{1}{q(1-q)}.  \label{eq:sd} 
 \end{eqnarray}
Conditionally on the value of OR, we can express variance ($\sigma^2$) as a function of two variables ($w$ and $p$) to emphasize that the standard deviation ($\sigma$) will vary depending on the study design and population prevalence of exposure among cases (we note that $q$ can be expressed in terms of $p$ and OR as $q = p/\left[ (1-p)\OR + p\right]$).
Alternatively, conditionally on the observed OR, one can express $\sigma$ in terms of the exposure probability,  $v=\Pr(E)$, and risk of disease among exposed as:
\begin{eqnarray}
  \sigma^2 &=& \frac{1}{v} \frac{1}{\Pr(D|E)\left[1-\Pr(D|E)\right]} \label{sigma.v} \\  \nonumber
  &+& \frac{1}{1-v} 
      \frac{1}{\Pr(D|\bar{E})\left[1-\Pr(D|\bar{E})\right]}, 
\end{eqnarray}
with $\Pr(D|\bar{E})=1/(1-\OR\left[1-1/\Pr(D|E)\right])$.

To obtain maximum possible value of the standardized $\log(\OR)$, we first need to minimize $\sigma$, conditional on the OR value, with respect to its two parameters. For example, if we set the first partial derivative of Eq. (\ref{sigma.v}) with respect to $v$ to zero, and solve the resulting equation in terms of $\Pr(D|E)$, it follows that:  
\begin{eqnarray}
  v_m &=& \argmin_v \sigma = \frac{1}{1 + \RR \,\, \sqrt{\OR^{-1}} }, \label{vm}
\end{eqnarray}
where RR represents relative risk. Further, setting the first partial derivative of Eq. (\ref{sigma.v}) with respect to $\Pr(D|E)$ to zero and plugging in $v_m$ instead of $v$ results in 
\begin{equation}
  \Pr(D|E)= 1 - \frac{1}{1 + \sqrt{\OR}}, \label{eq5}
\end{equation}
and
\begin{equation}
\Pr(D|\bar{E}) = \frac{1}{1 + \sqrt{\OR}} = 1 - \Pr(D|E). \label{eq6}
\end{equation}
Now, substituting Eqs (\ref{eq5}) and (\ref{eq6}) into Eq. (\ref{vm}), we obtain  $v = \nicefrac{1}{2}$. Similarly, operating with Eq. (\ref{eq:sd}), we can express the minimum $w$ value in terms of $p$ as
\begin{eqnarray}
   w_{m} &=& \argmin_w \sigma = \frac{1}{1 + \frac{p}{q} \,\, \sqrt{\OR^{-1}} }, \label{wm}
\end{eqnarray}
where $q = p/\left[ (1-p)\OR + p\right]$. Then, we can obtain an equivalent expression for $w$ as just we did for $v$, $w =  \nicefrac{1}{2}$.
Using the conditional value of $\sigma$, the maximum standardized log(OR) is:
\begin{eqnarray}
   \gamma &=& \frac{ \log(\OR) }{ 2 \sqrt{2 + \frac{1+\OR}{\sqrt{\OR}}} }. \label{defgamma}
\end{eqnarray}
Using the identity $\frac{1+\OR}{\sqrt{\OR}} = 2 \cosh\left( \frac{\log(\OR)}{2}\right)$,
\begin{eqnarray}
   \gamma  &=& \frac{\log (\OR)}{2 \sqrt{2} \sqrt{1+\cosh \left(\frac{\log (\OR)}{2}\right)}} \\
 &=&  \frac{\log (\OR)}{4 \cosh \left(\frac{\log (\OR)}{4}\right)}.
\label{max.gamma}
\end{eqnarray}
Equation (\ref{max.gamma}) depends only on the logarithm of odds ratio, but it is not monotone in it: $\gamma$ reaches its maximum for log(OR) value at about 4.7987..., 
\begin{equation*}
\gamma_{\max} (\log(\OR) = 4.7987...) = 0.6627...
\end{equation*}
Surprisingly, as log(OR) exceeds that value, the corresponding normalized coefficient, $\gamma$, starts to decrease. Further, although Equation (\ref{max.gamma}) depends only on log(OR), its maximum can only be attained at the specific values of population parameters. Namely, (a) $v=w=  \nicefrac{1}{2}$, (b) $\Pr(D|\bar{E}) = 1 - \Pr(D|E)$ from Eq.(\ref{eq6}), which implies $\RR^2 = \OR$, and (c) $\log(\OR) = 4.7987...$, which implies log(OR) = 121.354$\dots$ and $\Pr(D|E) = \Pr(E|D) = 0.9167782798\dots$

Finally, so far we relied on the assumption that epidemiological data was summarized by a 2$\times$2 table. To investigate whether the standardized log(OR) varies between -0.6627... and 0.6627... in the logistic regression setting with continuous predictors or with additional covariates, we turned to simulation experiments. For a continuous predictor $X$, we assumed that $X \mid Y = 1 \sim N(\mu_1, s)$ and denoted its probability mass function (PDF) $\psi_{\mu_1}$; similarly, we assumed $X \mid Y = 0 \sim N(\mu_0, s)$ with the PDF $\psi_{\mu_0}$. Then, according to the Bayes rule:
\begin{equation}
  \Pr(Y = 1 \mid X = x) = \frac{\psi_{\mu_1} \Pr(Y = 1)}{\psi_{\mu_1} \Pr(Y = 1) + \psi_{\mu_0} \Pr(Y = 0)}.
  \label{logist1}
\end{equation}
Setting $\Pr(Y = 1) = w$ and $\Pr(Y = 0) = 1- w$, we can re-write Eq. (\ref{logist1}) as:
\begin{equation}
  \Pr(Y = 1 \mid X = x) = \frac{1}{1 + \frac{\psi_{\mu_0}(1-w)}{\psi_{\mu_1}w}},
  \label{logist2}
\end{equation}
or equivalently obtain the logitic regression model:
\begin{equation}
  \Pr(Y = 1 \mid X = x) = \frac{1}{1 + \exp\left[-(\alpha + \beta x)\right]},
  \label{logist3}
\end{equation}
where $\alpha = -\left\{\log(\frac{1-w}{w}) + \beta \frac{\mu_0 + \mu_1}{2}\right\}$ and $\beta = \left(\frac{\mu_1-\mu_0}{s^2}\right)$ \cite{cornfield1962joint}. According to this model, we generated a continuous predictor from a mixture of two normal distributions. A binary predictor was generated from a Bernoulli distribution with a random success probability. Logistic parameters (intercept and slope for the predictor $X$) were generated by randomly sampling $p$ and $q$ from a uniform distribution and then setting $\alpha = \log(q/(1-q))$ and $\beta = \log(p/(1-p)) - \alpha$. Other covariates and their coefficients were randomly generated from a normal distribution. Across simulations, we investigated the range of $\beta / \sqrt{\Var(\beta) \times N}$ values and confirmed that it is bounded from -0.6627... to 0.6627... 

\subsection{Connection to the Laplace Limit Constant}
\begin{figure}[th!]
\centering
 \includegraphics[width=0.4\linewidth]{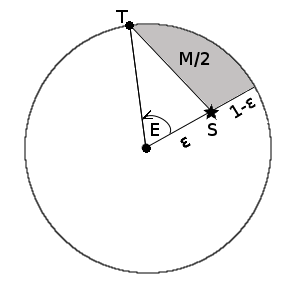} 
\caption{The Kepler equation: geometric interpretation. Given the knowledge of the area $M$ and the distance to the origin, $\varepsilon$, solve for the angle $E$ in $M = E - \varepsilon \sin(E)$ 
  .
}
\label{fig:KE}
\end{figure}

It turns out that there is an interesting connection between the expression for $\gamma_{\max}$ and the famous Kepler Equation (KE) for orbital mechanics, $M = E - \varepsilon \sin(E)$. Geometric interpretations of $M$, $E$ and $\varepsilon$ are illustrated by Figure \ref{fig:KE}. Specifically, suppose that one is inside a circular orbit, rescaled to be the unit circle, at the position S denoted by ``$\large{\star}$''. The shortest path to the orbit has length $1-\varepsilon$. A celestial body traveles the orbit from that point to point T. Given the area $M/2$ and distance $1-\varepsilon$, we want to determine the angle $E$. These three values are related to one another by Kepler's Equation. Planetary orbits are elliptical, so the actual orbit is along an ellipse inside of the unit circle.
Still, the calculation of the {\em eccentric anomaly}, $E$, is a crucial step in determining planet's coordinates along its elliptical orbit at various time points.

KE is transcendental, i.e., with no algebraic solution in terms of $M$ and $\varepsilon$, and it has been studied extensively since it is central to celestial mechanics. Colwell \cite{colwell1993solving} notes that ``in virtually every decade from 1650 to the present'' there have been papers devoted to the Kepler Equation in the book suitably named ``Solving Kepler's Equation over three centuries.'' The solution to KE 
involves the condition equivalent to Eq. (\ref{max.gamma}). Namely, the solution can be expressed as the power series in $\varepsilon$, provided $|\varepsilon \sin(E)| < |E-M|$ and that $\varepsilon < \psi / \cosh(\psi), \psi = |E-M|$, which is the ``Laplace Limit Constant,'' LLC \cite{plummer1918introductory}. The detailed mathematical derivation of the connection between Eq. (\ref{max.gamma}) and LLC is provided in ``Supplemental Materials (S-1).'' 

\subsection{The proposed normalized measure of effect size and its distribution} \label{sec:gamma}
\begin{figure}[th!]
\centering
 \includegraphics[width=0.4\linewidth]{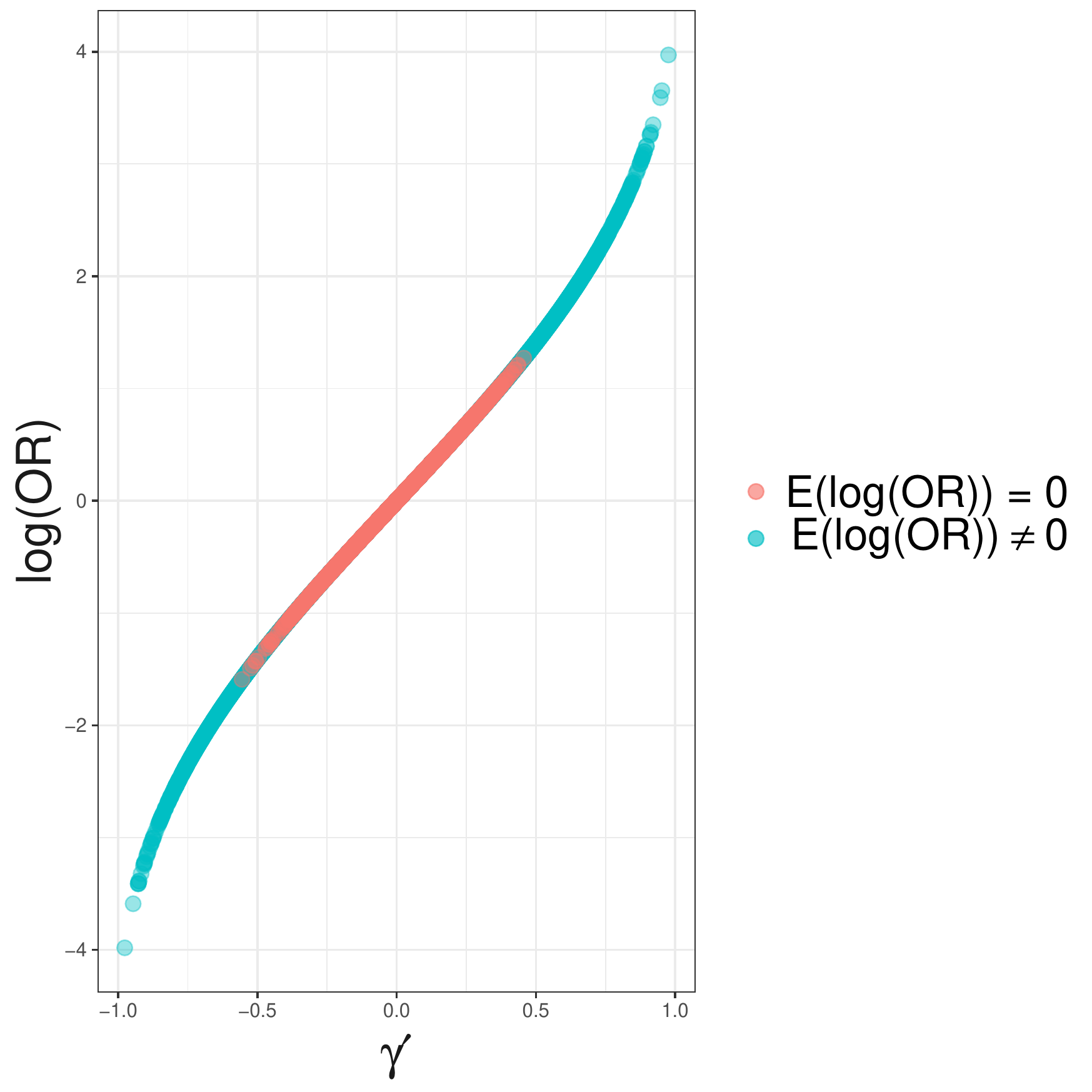} 
\caption{The relationship between log(OR) and $\gamma'$. The figure illustrates that under the null hypothesis (log(OR) $=$ 0), the relationship between sample values of log(OR) and $\gamma'$ is approximately linear, and under the alternative hypothesis (log(OR) $\neq$ 0) the relationship is monotone in the interval $\nicefrac{1}{121}\Nb <\Nb \OR\Nb <\Nb 121$.
}
\label{fig:logOR_gamma}
\end{figure}

As we showed above, at any value of $\log(\OR)$, the maximum of its standardized value is
\begin{eqnarray*}
   \gamma = \frac{ \log(\OR) }{ 2 \sqrt{2 + (1+\OR) / \sqrt{\OR}} }.
\end{eqnarray*}
The bounded nature of $\gamma$ (ranging between $-$LLC and LLC) suggests a new normalized measure of effect size, $\gamma^{\prime} = \gamma / \LLC$, that has the range $-1\Nb \leq \Nb\gamma^{\prime}\Nb \leq \Nb1$. The new statistic is appropriate as a measure of effect size within a very wide range of odds ratios, $\nicefrac{1}{121}\Nb <\Nb \OR\Nb <\Nb 121$, where it is monotone in $\OR$. For instance, Figure \ref{fig:logOR_gamma} shows that under the null hypothesis, the relationship between log(OR) and $\gamma'$ is almost linear, and under the alternative hypothesis, the relationship is close to linear and monotone, as long as $\nicefrac{1}{121}\Nb <\Nb \OR\Nb <\Nb 121$ (these are rounded to integer OR values  before the LLC maximum is reached).

Although $\gamma^{\prime}$ is derived by using the range of the standardized $\log(\OR)$, it is not a standardized measure in the same sense as scaling by standard deviation.  It is rather analogous to a coefficient denoted by $D^\prime$, which is commonly used in genetics to measure association between alleles at a pair of genetic loci (linkage disequilibrium, LD) \cite{lewontin1964interaction}. $D^\prime$ is akin to $\gamma^{\prime}$, because it is similarly obtained by taking a raw measure of LD and dividing it by its maximum value (which is a function of allele frequencies) to yield the $-1 \le D^\prime \le 1$ range.

Using the first order Taylor series approximation, we derive an asymptotic variance of $\gamma^\prime$, as well as one- and two-sided asymptotic test statistics, as follows:
\begin{eqnarray*}
   \Var\left(\widehat{\gamma^\prime}\right) &=& \hat{\sigma}^2 \left(\frac{\sech\left[\nicefrac{\log\left(\widehat{\OR}\right)}{4}\right] 
\left(4 - \log\left(\widehat{\OR}\right) \tanh \left[\nicefrac{\log\left(\widehat{\OR}\right)}{4}\right] \right) }{ 16 \times \LLC } \right)^2. \\
  T &=&  \sqrt{N} \frac{\widehat{\gamma^\prime}}{\sqrt{\Var\left(\widehat{\gamma^\prime}\right)}},   \\
\end{eqnarray*}
which simplifies to
\begin{eqnarray}
T  &=& \sqrt{N}\frac{4 \log\left(\widehat{\OR}\right)}{ \hat{\sigma} \,\,\left(4 - \log\left(\widehat{\OR}\right) \tanh\left[ \nicefrac{\log\left(\widehat{\OR}\right)}{4} \right]\right)}.
\label{eq:test}
\end{eqnarray}
The asymptotic distributions for one- and two-sided statistics are
\begin{eqnarray*}
T &\Asym& \text{Normal}(0,1), \\
T^2 &\Asym& \chi^2_{(1)},
\end{eqnarray*}
where $\hat{\sigma}$ is defined as before:
\begin{eqnarray*}
   \hat{\sigma} &=& \sqrt{\frac{1}{\hat{w}} \frac{1}{\hat{p}(1-\hat{p})} +
     \frac{1}{1-\hat{w}} \frac{1}{\hat{q}(1-\hat{q})}}.
\end{eqnarray*}
We show by simulation experiments that the null distribution of this new statistic reaches the asymptotic chi-square quicker than the commonly used $X^2\Nb =\Nb \log(\widehat{\OR})^2\Nb /\Nb \hat{\sigma}^2$ and that the new statistic provides higher power under the alternative hypothesis.

We note that two other well-known transformations of the OR with the range from -1 to 1 are Yule's coefficients: $\mathcal{Y} = \frac{\sqrt{\OR}-1}{\sqrt{\OR}+1}$, the coefficient of colligation, and $\mathcal{Q} = \frac{\OR-1}{\OR+1}$ \cite{yule1912methods}. Interestingly, using the identity $\frac{\sqrt{x}-1}{\sqrt{x}+1}=\tanh\left(\nicefrac{\log(x)}{4}\right)$, the statistic $T$ can be expressed as a function of $\mathcal{Y}$:
\begin{eqnarray}
   T  &=& \sqrt{N}\frac{4 \log\left(\widehat{\OR}\right)}{ \hat{\sigma} \,\,
      \left(\hat{\mathcal{Y}} - 4\right)}.
\label{eq:gamma_y}
\end{eqnarray} 
Further, note that $4 \arctanh\left(\mathcal{Y}\right) = 2 \arctanh\left(\mathcal{Q}\right) = \log(\OR)$. The $\arctanh$ transformation (to $\log(\OR))$, known as Fisher's variance stabilizing transformation \cite{fisher1915frequency}, is expected to improve the rate of asymptotic convergence to the normal distribution, thus we do not anticipate that the asymptotic test statistics based $\mathcal{Y}$ and $\mathcal{Q}$ would be competitive when compared to the $Z$ statistic based on the $\log(\OR)$. Nevertheless, we obtained approximate variances for $\mathcal{Y}$ and $\mathcal{Q}$ using the first order Taylor series approximation (the same type of approximation that yields the asymptotic variance for $\log(\OR)$) as follows:
\begin{eqnarray*}
   \widehat{\Var}\left(\widehat{\mathcal{Y}}\right) &=& \frac{1}{N}\left[\frac{\hat{p}}{\hat{w}(1-\hat{p}) \hat{q}^2 \left(\sqrt{\frac{\hat{p} (1-\hat{q})}{(1-\hat{p}) \hat{q}}}+1\right)^4}+\frac{1-\hat{q}}{(1-\hat{w}) (1-\hat{p})^2 \hat{q} \left(\sqrt{\frac{\hat{p} (1-\hat{q})}{(1-\hat{p}) \hat{q}}}+1\right)^4}\right], \\
   \widehat{\Var}\left(\widehat{\mathcal{Q}}\right) &=& \frac{1}{N}\left[\frac{4 (1-\hat{p}) \hat{p} (\hat{q}-1)^2 \hat{q}^2}{(1-\hat{w}) (\hat{p}+\hat{q}-2 \hat{p}\, \hat{q})^4}-\frac{4 (\hat{p}-1)^2 \hat{p}^2 (\hat{q}-1) \hat{q}}{\hat{w} (\hat{p}+\hat{q}-2 \hat{p}\, \hat{q})^4}\right].
\end{eqnarray*}
Our expressions turn out to be equivalent to squared standard errors for $\mathcal{Y}$ and $\mathcal{Q}$ as stated by Yule\cite{yule1912methods}, but his formulas are given in terms of four counts, while we separate frequencies from the reciprocal of the total sample size. Via simulations, we confirmed that the statistic for $\mathcal{Y}$ tends to be more conservative and less powerful than $Z$, while the statistic for $\mathcal{Q}$ is anti-conservative and reaches the nominal 5\% size only around $N=1,000$. However, these results are omitted here and we focus instead on comparisons of statistics based on $\gamma^{\prime}$ and $\log(\OR)$.

\subsection{Approximate Bayesian inference} \label{sec:bayes}
The rationale for using standardized coefficients (e.g., standardized log odds ratio) as measures of effect size in epidemiologic studies has been questioned and it has been suggested that standardized coefficients are insufficient summaries of effect size \cite{greenland1986fallacy, greenland1991standardized}. However, standardized effects can be utilized efficiently for delivering approximate Bayesian inference. Specifically, we propose to employ standardization as an intermediate step that yields posterior inference for parameters of interest (such as $\gamma^\prime$). The key to this approach is the observation that it is often straightforward to obtain an approximate posterior distribution for standardized effects ($\delta=\nicefrac{\mu}{\sigma}$) using a noncentral density as likelihood. Once such standardized posterior distribution is estimated, it can next be converted to an approximate posterior distribution for a parameter of interest, $\mu$. Our approach parallels semi-Bayes (also called partial-Bayes) methods, where explicit priors are used only for a subset of parameters \cite{greenland2006bayesian,greenland2007bayesian,greenland2009bayesian,wakefield2008ijepi}.

Let $\xi = \sqrt{N}  \times \delta$ denote the noncentrality parameter of the raw effect size density (for instance, $Z \sim N(\xi, 1)$ or $X^2 \sim \chi^2_1(\xi)$). To obtain an approximate posterior distribution, one needs to specify a prior distribution for a raw measure of effect size, $\mu$, as a binned frequency histogram, with a finite mixture of values $\mu_1, \mu_2, \ldots, \mu_B$ (the mid-values of bins) and the corresponding probabilities, $\Pr(\mu_i)$ (the height of bins as percent values). For example, if the effect size is measured by $\mu$=log(OR), such binned frequency histogram may be bell-shaped with a sizable spike around zero, indicating that the majority of risk effects are anticipated to be small. Alternatively, if the effect size is measured by log$^2$(OR), the frequency histogram may be L-shaped, with a spike of the mass again at about zero. 

Next, we employ an approximation to a fully Bayesian analysis (which would have required a joint prior distribution for both $\mu$ and $\sigma$), and ``dress'' the raw parameter, by plugging in the estimate of the standard deviation, to obtain values of $\delta_i = \nicefrac{\mu_i}{\hat{\sigma}}$ and $\xi_i = \sqrt{N} \delta_i$. Then, given the observed value of a test statistic $T=t$, the posterior distribution of the standardized effect size will also be a finite mixture, calculated as:
\begin{eqnarray}
   \Pr(\xi_j \mid T = t) = \frac{\Pr(\mu_j) f(T = t \mid \xi_j)}{\sum_{i=1}^B \Pr(\mu_i) f(T = t \mid \xi_i)}, \label{postd}
\end{eqnarray}
where $f$ is the test statistic density with the non-centrality parameter $\xi_i, \; i = 1, \ldots, B$. Once the posterior distribution for the standardized effect size (times $\sqrt{N}$), is evaluated, one can approximate the posterior distribution for the raw parameter of interest by ``undressing'' it, i.e., multiplying by the sample standard deviation and scaling by the square root of the sample size. For example, 
\begin{eqnarray}
   \Pr(\gamma'_i \mid T=t) = \Pr\left(\xi_i \cdot \sqrt{\widehat{\Var}\left(\widehat{\gamma^\prime}\right)}  / \sqrt{N} \mid T=t\right). \label{postd:undr}
\end{eqnarray}
From this approximate posterior distribution, one can then obtain an effect size estimator as the posterior mean by taking a weighted sum (e.g., $E(\gamma' \mid T=t) = \sum_{i=1}^B \gamma'_i \Pr(\gamma'_i \mid T = t)$), construct posterior credible intervals, etc. `Approximation' here refers to approximating a fully Bayesian modeling: our approach is a compromise between the frequentist and the Bayesian methodologies due to the usage of plug-in frequentist estimates for certain parameters. Although the posterior distribution for the raw effect size obtained via our method is approximate (due to plugging in a point sample estimate of the standard deviation), it is nevertheless highly accurate, as we demonstrate through our simulation experiments. 

\section{RESULTS}
\subsection{Simulations: Frequentist properties} \label{sec:pwr}
For an investigation of the statistical properties of the proposed procedures vis a vis traditional $Z$-tests, we now turn to simulation experiments. A supplemental materials appendix (S-2) provides details for the simulation setup. 

Our simulation experiments were not intended to include various types of existing statistics. Instead, we have explicitly focused on the two statistical measures described in our introduction in an `apples to apples' comparison -- that is, between two similarly derived Wald test statistics, both of which are based on the transformation of the odds ratio to measure effect size. The basic model for the OR estimates, without stratification or covariates, follows from the basic $2 \times 2$ contingency table. [We note that other investigations provide thorough coverage of the performance of alternative tests for contingency table associations, but have not covered the comparisons of interest to us \cite{larntz1978small,zaykin2008correlation}].

Type-I error rates for the two statistical measures, calculated under the null hypothesis of no effect, $\log(\OR)\Nb =\Nb 0$, are shown in Table 2. For small number of cases, the test based on $\log(\OR) = 0$ behaves conservatively, while the size of the test based on $\gamma^\prime$ is considerably closer to the nominal level of $\alpha = 0.05$. As the number of cases increases, the size of both tests approaches the nominal level. Table 3 shows statistical power of the two tests for the different combinations of $\log(\OR)$, its variance, and the number of cases. For all combinations of parameters considered, the $\gamma'$-based test has higher statistical power than the $Z$-test, particularly for small sample sizes. 

\begin{figure}[th!]
\centering
 \includegraphics[width=0.4\linewidth]{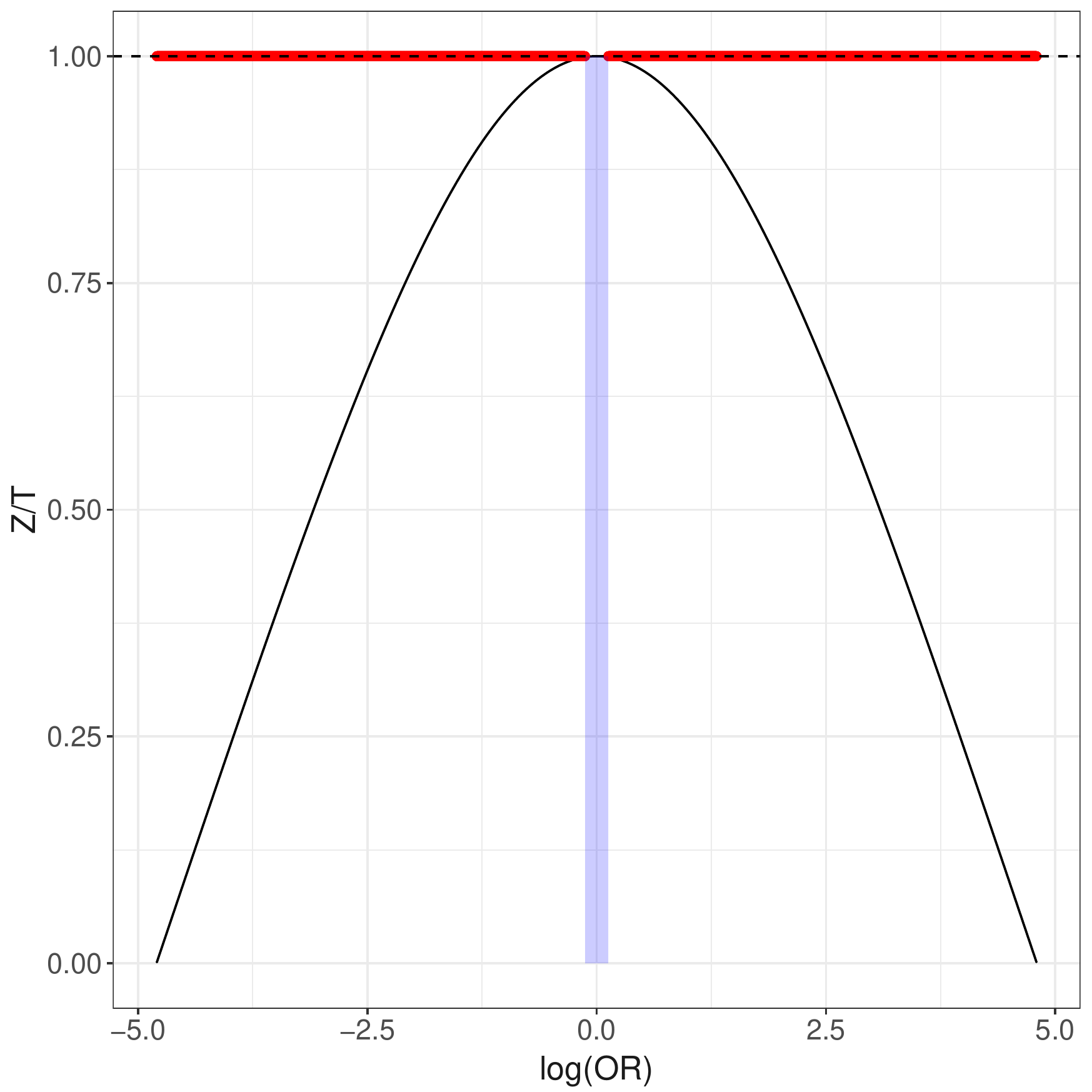} 
\caption{The range of $Z/T$-ratio values for $\nicefrac{1}{121}\Nb <\Nb \OR\Nb <\Nb 121$. The red line highlights log(OR) values, for which $\gamma'$-based $T$-statistic considerably exceeds $Z$-statistic. The blue rectangular highlights log(OR) values near the null hypothesis, for which the two statistics are similar to one another. Note that for all values of log(OR), $Z$-value never exceeds $T$-value.
}
\label{fig:r}
\end{figure}

We further note that the power of these two-sided tests can be investigated analytically by plugging in the population parameters, $p,q,w$ and considering the ratio of $Z$ and $T$ values, $Z/T$. Sample size and variance cancel out and their ratio becomes only a function of log(OR):
\begin{eqnarray*}
  Z/T = \frac{4 - \log(\OR) \tanh \left[ \nicefrac{\log\left(OR \right)}{4} \right] }{4}.
\end{eqnarray*}
Figure \ref{fig:r} illustrates that for all odds ratio values within the $(\nicefrac{1}{121} - 121)$ interval, $T$-statistics are at least as large as $Z$-statistics. Under the null (true log(OR) = 0), the ratio $Z/T$ is one, and the two statistics are equivalent. 

\subsection{Simulations: Bayesian properties}
Averaged across simulations, Table 4 reports (a) the true mean value of the raw parameter, $E(\gamma')$, corresponding to the maximum observed test statistic; (b) the posterior expectation  $E(\gamma^{\prime}\Nb\mid\Nb Z_{\max})$; (c) the average for the frequentist estimator of gamma prime, $E(\widehat{\gamma'})$; and (d) the average probability to contain the true log(OR) value by the high posterior density interval. 

On the one hand, Table 4 shows that posterior expectation is very close to the true average effect size value, even when posterior inference was performed using small sample sizes and extreme selection (i.e., the top-ranking result taken out of one million statistical tests). On the other hand, Table 4 illustrates that the frequentist estimator of $\gamma'$ is subject to the winner's curse and grossly exaggerates the true magnitude of effect size. Finally, after comparing posterior convergence to the corresponding nominal levels, it is clear that the posterior interval's performance is satisfactory. 

\subsection{Co-occurance of cannabis-use associated SEPEs }

To investigate patterns of SEPE-SEPE co-occurance, we obtained unweighted counts for pair-wise SEPEs and organized them using $2 \times 2$ tables, in which individuals were classified as reporting or not SEPE$_i$, SEPE$_j$, for all $j > i, i=1, \ldots, 17$ with the exception of 1-10, 2-11, 3-4, 5-15, 7-14, 13-15, 13-17, and 15-17 pairs. These pairs were excluded because the answer to the $j$th SEPE was conditional on the positive answer to the $i$th SEPE. For example, it is impossible for one to be `unable to cut down or stop' using cannabis (SEPE$_{10}$), unless the one is `wanting or trying to cut down or stop' (SEPE$_1$). These data manipulations resulted in 128 (17 choose 2 minus 8) $2 \times 2$ tables calculated for each year across the ten years under study, 2004 - 2014.

Next, we focused on the year 2004. In a correction step, we added 1/2 to each cell count and calculated 128 ORs for SEPE$_i$-SEPE$_j$ co-occurrence.  The addition of 1/2 to cell counts is known as the Haldane-Anscombe (or `pseudo-Bayes') correction, and is commonly used to improve asymptotic convergence to normality of the test statistic for log(OR) \cite{haldane1956estimation,anscombe1956estimating,lawson2004small,agresti1999logit}. Given the calculated ORs (or log(OR)s), we used the approximate Bayesian inference to obtain the posterior distribution for the standardized effect size (Eq. \ref{postd}).  We assumed that the prior distribution of log(OR) is formed by the 50:50 mixture of the two truncated normal distributions: log(OR)$\mid H_0 \sim \text{Truncated Normal}(0, 0.001)$ with the truncation parameter set to 0.01, and log(OR)$\mid H_1 \sim\text{Truncated Normal}(0, W = 0.67)$ with the truncation parameter set to 4.8. That is, \textit{a priori} we assumed that there is a 50\% chance that the observed association is false (i.e., log(OR) comes from the zero-centered normal distribution with a sizable spike at zero and no values greater in absolute value than 0.01), and 50\% chance that it is real. The standard deviation $W$ was chosen so that there is a 5\% \textit{a priori} chance of encountering OR$\geq 3$, and 5\% chance of encountering OR$\leq 1/3$; the truncation parameter was set to 4.8, which correponds to the maximum OR of about 121. Next, we sampled 7 million values from the prior mixture and ``dressed'' prior sampled values by multiplying them by the observed $\sqrt{N_k} / \sqrt{\Var(\log(\text{OR}_k))}, k = 1, \ldots, 128$, thus obtaining 128 discretized bins for the prior distribution of the non-centrality parameter $\xi$. Finally, using Eq. (\ref{postd}), we obtained 128 posterior densities for the non-centrality parameter for the year 2004.

Next, we used 2004 posterior densities as prior densities for the year 2005. Then, with the 2005-2014 data, we iteratively repeated these steps until we obtained posterior non-centrality densities through to the year 2014. In a final step, we ``undressed'' the posterior noncentralities (Eq. \ref{postd:undr}) and calculated posterior expectations, with the corresponding 95\% credible intervals, for 128 $\gamma^\prime$s.

Table \ref{tab:sepe} provides a summary of what we found about co-occurrence of SEPE-SEPE pairs, with posterior expectations of scaled from $\gamma^\prime$ -1 to 1.0, and with a focus on co-occurrences observed within $\sim$90 days after first cannabis use. The first row of Table \ref{tab:sepe} is focused on SEPE$_1$ and its degree of co-occurrence with SEPE$_j$ as $j = 2, 3, \ldots, 9, 11, \ldots, 17$. Positive values of $\gamma^\prime$ indicate an elevated co-occurrence of the SEPE-SEPE pairs. When the associated 95\% Bayesian credible intervals cover zero, the degree of association linking SEPE$_1$ with SEPE$_j$ must be regarded as weak. Based on the results in the final row of Table \ref{tab:sepe}, we might conclude that once new cannabis initiates want or try `to cut down or stop using cannabis,' they also are more likely to have experienced SEPE$_3$, SEPE$_5$, SEPE$_7$, or SEPE$_9$. As such, this subset of the SEPE-SEPE pairs might qualify as emergent therapeutic targets in a process of becoming a case of cannabis use disorder.

The second row displays corresponding estimates for the co-occurrence of $SEPE_2$  relative to  SEPE$_j$ for $j > 2$.  Here, we note that when new cannabis initiates are `spending a lot of time getting or using cannabis,' the survey data indicate statistically robust estimates of $\gamma^\prime$ for all of the SEPE$_3$-SEPE$_{17}$ pairs, and an especially sizeable $\gamma^\prime$ estimate of 0.75 for SEPE$_{14}$ (i.e., continuing to use cannabis despite problems with family or friends; credible interval = 0.71, 0.90).

The remaining rows of Table \ref{tab:sepe} can be read out in analogously. The result is a view of the `greater-than-chance' co-occurrence of SEPE, arranged in pairs that can be used to evaluate potential syndromic or pre-syndromic therapeutic targets for investigators interested in prevescalation or disruption of the potentially pathological processes that lead toward or beyond this stage of formation of cannabis syndromes.

\begin{figure}[th!]
\centering
 \includegraphics[width=1\linewidth]{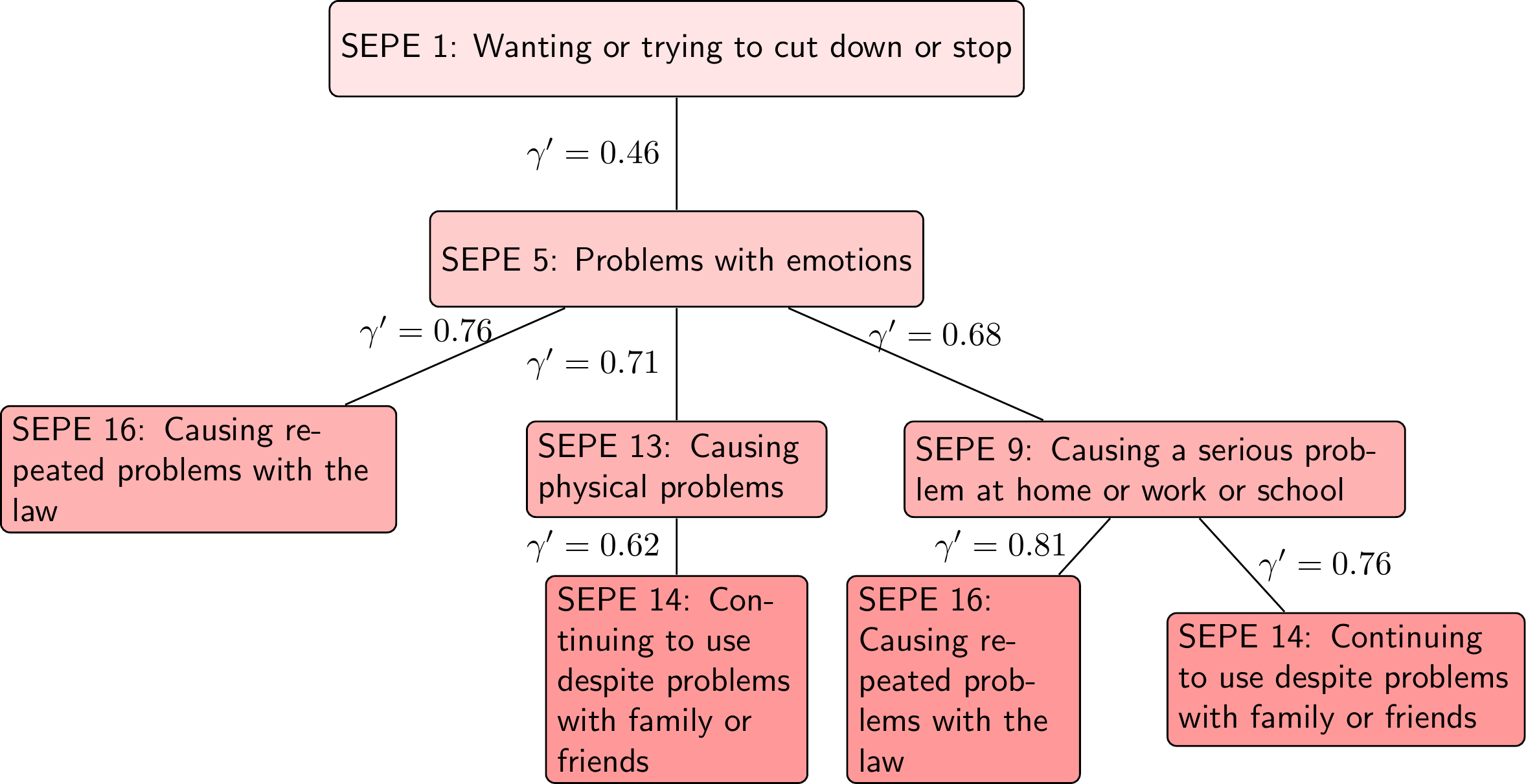} 
\caption{One of the likely progressions of the combinations of SEPEs within the first 90 days after cannabis use onset (2004-2014 NSDUH).
}
\label{fig:sepe}
\end{figure}

In a forecast of future prospective and longitudinal research that builds from this cross-sectional view of SEPE-SEPE pairs, one might posit a possibly accelerated CUD development process. For example, consider Figure \ref{fig:sepe} and its depiction of a plausible chain of progression from SEPE$_1$, which was found to have the largest SEPE attack rate estimate in these data on newly incident cannabis users. Table \ref{tab:sepe} shows that the largest $\gamma^\prime$ seen for the vector of SEPE$_1$-SEPE$_j$ pairs is the estimate for $j=5$. One possibility is that there are cannabis-attributable `problems with emotions,' once initiates want or try to cut down or stop using cannabis. Alternatively, once cannabis-attributable problems with emotions appear, Table  \ref{tab:sepe} provides empirical guidance in the direction of larger values of $\gamma^\prime$ among SEPE$_5$-SEPE$_j$, $j = 6, \ldots, 17$ pairs, from which we might identify a likely set of cannabis SEPE. Prominent in this set might turn out to be cannabis causing a serious problem at home, work, or school (SEPE$_9$, $\gamma^\prime$ = 0.68), cannabis causing physical problems (SEPE$_{13}$, $\gamma^\prime$ = 0.71), or cannabis-caused repeated legal problems (SEPE$_{16}$, $\gamma^\prime$ = 0.76). 

In our presentation of these findings, we must acknowledge that the NSDUH data are not sufficiently fine-grained concerning the dimension of temporal sequencing. The estimates of $\gamma^\prime$ that are presented in this paper's tables are not time-sequenced, as they might be in prospective and longitudinal research on the addictive processes that lead relatively quickly from first cannabis use until the formation of early emerging cannabis dependence syndromes. Nevertheless, these cross-sectional estimates resemble what we can see when patients infected with a novel virus then experience fever, headache, myalgia, dry cough, and other manifestations of the virus-influenced pathological process, and the patients are not seen until they qualify for testing based on the symptoms and signs they have experienced before being tested. Here, we have no virus infection, but we have the first exposure to a cannabis product as an agent that might or might not be followed by a pathological process that leads eventually toward a cannabis use disorder. These new cannabis initiates are observed with a relatively short interval of time after cannabis onset. As might be the case after a virus infection, there is no time-sequencing data about which came first, the fever or the headache. Rather, in the early stages of investigation of post-exposure syndromes of this type, we must rely upon cross-sectionally gathered evidence to guide later prospective and longitudinal studies to throw light on which features of the syndrome might come first and which might come later.

\section{DISCUSSION}
In this article, our primary intent has been to introduce `gamma prime' ($\gamma^\prime$) as a transformed odds ratio for use as a statistical measure of the magnitude of association as might link two binary response variables (here, SEPE) or to link a binary exposure variable (X) with a binary response variable (Y). In addition to our description of statistical properties and performance of $\gamma^\prime$ relative to its traditional odds ratio alternative, we have illustrated contemporary public health utility of $\gamma^\prime$ in novel research on what newly incident cannabis users experience within the first $\sim$90 days after cannabis onset.

Studying cannabis side effect problems and experiences (SEPE) assessed within the first $\sim$90 days after first cannabis use, we illustrate how the $\gamma^\prime$ statistic can be used to quantify the degree of association between pair of SEPE. Several implications can be seen. First, new evidence about SEPE-SEPE pairs suggests potential therapeutic target profiles of interest to intervention researchers who might have in mind novel medications or non-drug interventions intended to disrupt the process of becoming a case of cannabis use disorder. One result of public health significance can be `prevescalation' (i.e., preventing escalation) during the processes that develop after one starts to use cannabis, but before one qualifies as a formally diagnosable case of cannabis use disorder. The resulting overview of the natural history of an early progression to CUD can help clinicians and researchers who are trying to identify individuals with relatively non-toxic cannabis use experiences and to discriminate these individuals from users on an experience-trajectory that might require intensive clinical interventions.

The simulation studies described in this paper show that the $\gamma^\prime$ test statistic can provide better control of Type I error relative to the traditional $Z$-test when the sample size is constrained. Also, the statistical power of $\gamma^\prime$ seems to be at least as good as that of the traditional $Z$-test for null odds ratios ($=1$). In addition, we offer a simple and efficient approach for obtaining an approximate posterior distribution for $\gamma^\prime$ and demonstrate its robustness to selection bias, a feature that should promote the reliability of reported findings.

The new measure $\gamma^\prime$ is normalized by the Laplace Limit Constant to possess a range that runs from $-1.0$ to $1.0$. As a result, the non-negative positive values of $\gamma^\prime$ are indicative of co-occurrences, and the negative values of $\gamma^\prime$ are indicative of inverse co-occurrences. In applications of $\gamma^\prime$ to cause-effect and protective-effect associations, a robust positive sign on $\gamma^\prime$ might lead to an inference of causal influence. A robust negative sign on $\gamma^\prime$ might lead to an inference of a protective effect.

In this paper's subject matter context, the estimates for $\gamma^\prime$ provide evidence that many early-observed SEPE are co-occurring with other SEPE. We note that all estimates of the upper bounds for $\gamma^\prime$ have positive signs. An implication might be that no SEPE is serving to dampen the occurrence of the other SEPE.

Several limitations deserve attention, in addition to our previously mentioned caution about cross-sectional and retrospective survey estimates, which might not convey the same estimates to be found in prospective and longitudinal research. Nevertheless, it is not possible to design and conduct informative prospective and longitudinal research in epidemiology before we have starting estimates from cross-sectional studies (e.g., to provide information about statistical power projections).

We also note that there are no logistically feasible biological assays for the cannabis side effect problems and experiences we are studying in this work. In theory, pharmacological tolerance and withdrawal might be studied without a reliance upon subjectively felt experiences, but no one has accomplished this measurement task with epidemiological samples as large as the NSDUH samples. In consequence, the self-report ACASI assessments might be as good as it gets in large sample epidemiological field surveys.

We should express some uncertainty about whether $\gamma^\prime$ estimates for drugs other than cannabis will serve well. For opioid drugs such as fentanyl and heroin, there are `left-hand-side' time-to-event errors (e.g., truncation), such that a newly incident user might die of an overdose within a $\sim$90 day interval after first drug use. This type of time-to-event error most likely is constrained in research on cannabis and other drugs for which user-fatality rates from overdose occur rarely if at all.

In the domain of statistical limitations, we should note that $\gamma'$ should not be regarded as an approximate Pearson correlation coefficient ($\rho_{X,Y}$), nor does it behave as a standardized measure of effect size. That said, the usual standardized log(OR) can be approximately related to the standardized slope and to the correlation coefficient  $\rho_{X,Y}\Nb =\Nb \beta\Nb\times\Nb (\sigma_X\Nb /\Nb \sigma_Y)$ in simple linear regression models. When both $X$ and $Y$ are binary, $\rho_{X,Y}$ can be expressed as:
\begin{eqnarray*}
   \rho_{X,Y} &=& (p - q) \frac{ \sqrt{w(1-w)} }{ \sqrt{v(1-v)} } \\ 
     & \approx & \ln(\OR) \sqrt{v(1-v)}\sqrt{w(1-w)}, \quad \,\,\,\, (\text{because} \,\,\, p-q \approx \ln(\OR)(1-v)v), \\ 
     & \approx & \delta.
\end{eqnarray*}
In addition, we should note that standardized coefficients may be used in practice as a ``scale-free'' measure. Nevertheless, it has been suggested that the magnitude of these statistical measures may not appropriately reflect relative importance of explanatory variables \cite{greenland1986fallacy, greenland1991standardized}. Given that our $\gamma'$ is not obtained using a regular standardization technique (i.e., scaling by the standard deviation), we omit arguments for and against the use of standardized coefficients in statistical practice.

In this work, we have provided an example of how the standardized logarithm of the odds ratio ($\delta$) can be used as a middle step towards an approximate posterior inference for a raw (non-standardized) parameter of interest. Based on an assumption that the prior distribution for the raw parameter is known precisely, we checked the performance of our method in terms of its resistances to the `winner’s curse' as well as robustness of estimation in the presence of multiple testing.

In this context, exact knowledge of the prior distribution is improbable and outside the boundaries of a research team’s reach. Nevertheless, assumptions about the prior distribution can be useful for the purpose of checking accuracy of methods performance in an ideal scenario. Assuming that the prior is known, proper posterior estimates should not overstate the effect size when the top-ranking associations are selected out of a large number of results. As for practical implementations, although the exact prior distribution may not be known, it can be specified realistically.

We recognize that the problem of a reasonable prior choice can be challenging. We also note that this problem is not unique to our proposed method and it is a problem that is ubiquitous within the Bayesian framework. Furthermore, in our current application, the $50:50$ mixture of two truncated normal distributions had little impact on our final reported results -- due to the sequential prior-poster update that allowed us to combine 2004-2014 results in a `meta-analytical fashion.'

\clearpage
\bibliographystyle{apacite}
\bibliography{Kepler}

\clearpage
\section*{Tables}
\begin{table}[th!]
  \begin{threeparttable}
    \caption{Estimated analysis-weighted incidence of cannabis SEPE soon after cannabis onset ($\sim$1-90 Days). Data from non‐institutionalized civilian 12‐to‐29‐year‐olds in the United States based on the National Surveys on Drug Use and Health, 2004 -- 2014 ($n = 3,710$ newly incident cannabis users).
    }
    \centering
    \begin{tabular}[th!]{lll} \hline
      SEPE & Description & Incidence Rate (95\% CI) \\ \hline
      1 & Wanting or trying to cut down or stop using cannabis & 0.54 (0.47, 0.61) \\
      2 & Spending a lot of time getting or using cannabis & 0.25 (0.20, 0.31) \\
      3 & Using the same amount but it has less effect & 0.15 (0.12, 0.19) \\
      4 & Needing more cannabis to get the same effect & 0.12 (0.96, 0.15) \\
      5 & Cannabis causing problems with emotions & 0.09 (0.07, 0.12) \\
      6 & Spending less time doing important activities & 0.09 (0.07, 0.12) \\
      7 & Causing problems with family members or friends & 0.09 (0.07, 0.12) \\
      8 & Being unable to keep limits (continuing to  use) &  0.09 (0.06, 0.13) \\
      9 & Causing a serious problem at home or work or school & 0.07 (0.05, 0.10) \\
      10 & Being unable to cut down or stop & 0.07 (0.05, 0.09) \\
      11 & Spending a lot of time getting over effects & 0.06 (0.04, 0.09) \\
      12 & Doing dangerous activities & 0.06 (0.04, 0.09) \\
      13 & Cannabis causing physical problems & 0.04 (0.03, 0.06) \\
      14 & Continuing to use despite problems with family or friends & 0.04 (0.03, 0.05) \\
      15 & Continuing to use despite emotional problems & 0.04 (0.02, 0.06) \\
      16 & Causing repeated problems with the law & 0.02 (0.01, 0.03) \\
      17 & Continuing to use despite physical problems & 0.01 (0.01, 0.02) \\ \hline
    \end{tabular}
    \begin{tablenotes}
    \item[] \hspace{-0.5 cm} Abbreviations: SEPE, side effect problem-experiences; CI, confidence intercal. 
    \item[] \hspace{-0.5 cm} \textit{Note}: Alcover et al. (2019) recently provided a description of the unweighted sample characteristics, which is readily accessible via this URL: \url{https://onlinelibrary.wiley.com/doi/abs/10.1111/ajad.12943}. For this reason, we direct the reader’s attention to this online resource in lieu of adding the table of unweighted sample characteristics to this journal article submission.
    \end{tablenotes}
  \end{threeparttable}
  \label{sepe}
\end{table}

\clearpage
\begin{table}[th!]
  \caption{The type-I error rate by the number of cases ($\log \OR = 0$).}
  \centering
  \begin{tabular}[th!]{lcc} \hline
    & $\log(\OR)$ & $\gamma'$ \\ \cline{2-3}
    $n_D=25$   & 0.029 & 0.051 \\ 
    $n_D=50$   & 0.038 & 0.050 \\
    $n_D=100$  & 0.043 & 0.049 \\
    $n_D=250$  & 0.046 & 0.049 \\
    $n_D=500$  & 0.048 & 0.049 \\
    $n_D=1,000$ & 0.048 & 0.049 \\  
    $n_D=5,000$ & 0.050 & 0.050 \\ \hline
  \end{tabular}
  \begin{threeparttable}
    \begin{tablenotes}
    \item[]  Abbreviations: OR, odds ratio.
    \item[] \textit{Note}: For small number of cases, the test based on $\log(\OR)$ behaves conservatively, while the size of the test based on $\gamma^\prime$ is considerably closer to the nominal level of $\alpha = 0.05$.
    \end{tablenotes}
     \label{type1}
  \end{threeparttable}
\end{table}

\clearpage
\begin{table}[th!]
  \caption{Power of the two tests by different levels of $\log(\OR)$ and $\tau$, assuming that $\log(\OR)\Nb \sim\Nb N(0,\tau)$.}
  \centering
  \begin{tabular}[th!]{lcccccccc} \\ \hline
    & $\log(\OR)$ & $\gamma'$ & $\log(\OR)$ & $\gamma'$ & $\log(\OR)$ & $\gamma'$ & $\log(\OR)$ & $\gamma'$ \\ \hline
    $\log(\OR)$$\sim$$N(0,\tau)$ $\Rightarrow$ & \multicolumn{2}{c}{$\tau=\frac{\log(2)}{\Phi^{-1}(1-0.05)} \approx$  0.42} & \multicolumn{2}{c}{$\tau$ = 0.5} & \multicolumn{2}{c}{$\tau$ = 1} & \multicolumn{2}{c}{$\tau$ = 2} \\ \cline{1-9}
    $n_D=25$    & 0.065 & 0.098 &  0.080 & 0.116 & 0.212 & 0.263  & 0.441 & 0.493 \\
    $n_D=50$    & 0.121 & 0.142 &  0.151 & 0.174 & 0.358 & 0.385  & 0.602 & 0.624 \\
    $n_D=100$   & 0.204 & 0.217 &  0.253 & 0.266 & 0.503 & 0.516  & 0.718 & 0.726 \\
    $n_D=250$   & 0.360 & 0.365 &  0.423 & 0.429 & 0.664 & 0.668  & 0.821 & 0.823 \\
    $n_D=500$   & 0.490 & 0.493 &  0.553 & 0.556 & 0.757 & 0.758  & 0.873 & 0.874 \\
    $n_D=1,000$  & 0.613 & 0.614 &  0.666 & 0.667 & 0.825 & 0.826  & 0.909 & 0.910 \\
    $n_D=5,000$  & 0.814 & 0.814 &  0.843 & 0.843 & 0.921 & 0.921  & 0.960 & 0.960 \\ 
    \hline
    Fixed OR $\Rightarrow$ & \multicolumn{2}{c}{OR = 1.25} & \multicolumn{2}{c}{OR = 2} & \multicolumn{2}{c}{OR = 3} & \multicolumn{2}{c}{OR = 4} \\ \cline{1-9}
    $n_D=25$   & 0.038 & 0.064 & 0.124 & 0.176 & 0.276 & 0.354 & 0.411 & 0.499 \\
    $n_D=50$   & 0.061 & 0.076 & 0.267 & 0.303 & 0.559 & 0.602 & 0.735 & 0.770 \\
    $n_D=100$  & 0.091 & 0.101 & 0.499 & 0.521 & 0.840 & 0.854 & 0.938 & 0.945 \\
    $n_D=250$  & 0.174 & 0.180 & 0.850 & 0.856 & 0.985 & 0.986 & 0.999 & 0.999 \\
    $n_D=500$  & 0.306 & 0.310 & 0.971 & 0.972 & 0.999 & 0.999 & 1     & 1     \\
    $n_D=1,000$ & 0.532 & 0.534 & 0.998 & 0.998 & 1     & 1     & 1     & 1     \\ 
    $n_D=5,000$ & 0.971 & 0.971 & 1 & 1 & 1     & 1     & 1     & 1     \\ 
    \hline
  \end{tabular}
  \begin{threeparttable}
    \begin{tablenotes}
    \item[]  Abbreviations: OR, odds ratio.
    \item[] \textit{Note}:  For all combinations of parameters considered, the $\gamma'$-based test has higher statistical power than the $Z$-test, particularly for small sample sizes.
    \end{tablenotes}
  \end{threeparttable}
\end{table}

\clearpage
\begin{table}[!th]
   \caption{Average true value, $E(\gamma')$, average posterior estimator, $E(\gamma' \mid  Z_{\max})$, and average frequentist estimator, $E(\widehat{\gamma'})$, assuming $\gamma'\sim N(0, \tau = 0.42)$ for the top-ranking (maximum) observed statistic ($Z$) selected out of $L$ tests. Averages refer to the mean value taken across simulation experiments.}
  \centering
  \begin{tabular}{llcccc}
    \hline
    \# of tests ($L$) & $n$ & $E(\gamma')$ & $E(\gamma' \mid  Z_{\max})$ & $E(\widehat{\gamma'})$ & Posterior coverage \\ \hline
    10,000   & 500   &  0.46 & 0.46 & 0.55 & 94\% \\
             & 750   &  0.47 & 0.47 & 0.53 & 95\% \\
             & 1,000 &  0.48 & 0.48 & 0.52 & 95\% \\
             & 1,500 &  0.48 & 0.48 & 0.51 & 95\% \\ \hline
   100,000   & 500   &  0.52 & 0.53 & 0.62 & 93\% \\ 
             & 750   &  0.53 & 0.54 & 0.60 & 95\% \\
             & 1,000 &  0.54 & 0.55 & 0.60 & 95\% \\
             & 1,500 &  0.55 & 0.55 & 0.58 & 95\% \\ \hline 
   500,000   & 500   &  0.56 & 0.57 & 0.67 & 92\% \\
             & 750   &  0.56 & 0.57 & 0.64 & 94\% \\
             & 1,000 &  0.58 & 0.58 & 0.63 & 94\% \\
             & 1,500 &  0.59 & 0.59 & 0.63 & 95\% \\ \hline
  1,000,000   & 500  &  0.58 & 0.59 & 0.69 & 91\% \\
             & 750   &  0.58 & 0.59 & 0.66 & 93\% \\
             & 1,000 &  0.61 & 0.61 & 0.66 & 95\% \\ 
             & 1,500 &  0.61 & 0.61 & 0.65 & 95\% \\ \hline
  \end{tabular}
   \begin{threeparttable}
    \begin{tablenotes}
    \item[] \textit{Note}:  Posterior expectation is very close to the true average effect size value, even when posterior inference was performed using small sample sizes and extreme selection, while the frequentist estimator of $\gamma'$ is subject to the winner's curse and grossly exaggerates the true magnitude of effect size.    \end{tablenotes}
  \end{threeparttable}
\end{table}

\clearpage
\begin{landscape}
  \begin{table}[!th]
     \caption{Posterior estimates for the estimated degree of association for each of the 128 SEPE$_i$-SEP$_j$ pairs developed soon after cannabis use onset ( within 1-90 days), as quantified by the gamma prime ($\gamma'$) statistic. Data from the United States National Surveys on Drug Use and Health, 2004-2014.}
  \centering
  \resizebox{1.5\textwidth}{!}{
  \begin{tabular}{l| cccccccccccccccc} 
     & 2 & 3 & 4 & 5 & 6 & 7 & 8 & 9 & 10 & 11 & 12 & 13 & 14 & 15 & 16 & 17 \\ \hline \hline
    1 & 0.01 & 0.23 & 0.02 & 0.46 & 0.16 & 0.22 & 0.03 & 0.31 & NA & 0.17 & 0.01 & 0.06 & 0.01 & 0.21 & 0.09 & -0.04 \\ 
     & (-0.07, 0.10) & (0.04, 0.40) & (-0.09, 0.12) & (0.29, 0.59) & (-0.01, 0.32) & (0.02, 0.42) & (-0.16, 0.23) & (0.11, 0.48) &  & (-0.05, 0.39) & (-0.11, 0.13) & (-0.15, 0.26) & (-0.13, 0.15) & (-0.01, 0.42) & (-0.15, 0.32) & (-0.30, 0.22) \\\hline
    2 & & 0.35 & 0.50 & 0.30 & 0.61 & 0.53 & 0.59 & 0.46 & 0.33 & NA & 0.43 & 0.32 & 0.75 & 0.53 & 0.37 & 0.36 \\  
     & & (0.19, 0.49) & (0.38, 0.61) & (0.14, 0.44) & (0.48, 0.72) & (0.40, 0.64) & (0.40, 0.76) & (0.30, 0.59) & (0.12, 0.53) & & (0.27, 0.57) & (0.10, 0.52) & (0.71, 0.90) & (0.35, 0.68) & (0.09, 0.61) & (0.00, 0.65) \\ \hline
    3 & & & NA & 0.40 & 0.36 & 0.19 & 0.19 & 0.41 & 0.05 & 0.47 & 0.13 & 0.51 & 0.10 & 0.42 & 0.43 & 0.05 \\
     & & &    & (0.22, 0.56) & (0.17, 0.53) & (-0.001, 0.38) & (-0.10, 0.46) & (0.21, 0.58) & (-0.19, 0.29) & (0.25, 0.66) & (-0.08, 0.35) & (0.29, 0.71) & (-0.15, 0.35) & (0.17, 0.62) & (0.18, 0.65) & (-0.27, 0.36) \\ \hline
    4 & & & & 0.32 & 0.37 & 0.38 & 0.62 & 0.44 & 0.50 & 0.30 & 0.39 & 0.44 & 0.54 & 0.42 & 0.43 & 0.58 \\
     & & & & (0.14, 0.48) & (0.21, 0.52) & (0.22, 0.52) & (0.45, 0.79) & (0.27, 0.58) & (0.31, 0.67) & (0.00, 0.57) & (0.22, 0.54) & (0.24, 0.61) & (0.37, 0.69) & (0.21, 0.59) & (0.20, 0.62) & (0.34, 0.79) \\ \hline
    5 & & & & & 0.55 & 0.49 & 0.19 & 0.68 & 0.03 & 0.59 & 0.52 & 0.71 & 0.44 & NA & 0.76 & 0.33 \\
     & & & & & (0.39, 0.68) & (0.34, 0.63) & (-0.09, 0.46) & (0.56, 0.91) & (-0.17, 0.22) & (0.39, 0.76) & (0.35, 0.66) & (0.63, 0.91) & (0.24, 0.62) & & (0.71, 0.91) & (-0.00, 0.62) \\ \hline
    6 &&&&&& 0.67 & 0.51 & 0.90 & 0.13 & 0.41 & 0.59 & 0.35 & 0.72 & 0.71 & 0.73 & 0.47 \\
     &&&&&& (0.55, 0.79) & (0.29, 0.70) & (0.89, 0.91) & (-0.11, 0.37) & (0.14, 0.63) & (0.44, 0.72) & (0.11, 0.57) & (0.66, 0.91) & (0.65, 0.91) & (0.68, 0.91) & (0.18, 0.71) \\ \hline
    7 &&&&&&& 0.45 & 0.68 & 0.26 & 0.15 & 0.61 & 0.52 & NA & 0.53 & 0.76 & 0.36 \\
     &&&&&&& (0.23, 0.64) & (0.58, 0.91) & (-0.00, 0.51) & (-0.13, .42) & (0.47, 0.73) & (0.33, 0.68) & & (0.33, 0.69) & (0.72, 0.91) & (-0.00, 0.67) \\ \hline
    8 &&&&&&&& 0.29 & 0.77 & 0.19 & 0.26 & 0.46 & 0.56 & 0.03 & 0.39 & 0.22 \\
     &&&&&&&& (-0.00, 0.55) & (0.74, 0.91) & (-0.15, 0.52) & (-0.01, 0.51) & (0.17, 0.69) & (0.33, 0.75) & (-0.22, 0.28) & (0.00, 0.70) & (-0.15, 0.55) \\ \hline
    9 &&&&&&&&& 0.41 & 0.62 & 0.61 & 0.55 & 0.78 & 0.73 & 0.83 & 0.56 \\
     &&&&&&&&& (0.17, 0.61) & (0.43, 0.83) & (0.45, 0.74) & (0.35, 0.72) & (0.76, 0.91) & (0.67, 0.91) & (0.81, 0.91) & (0.31, 0.79) \\ \hline
    10 &&&&&&&&&& 0.56 & 0.53 & 0.10 & 0.49 & 0.15 & 0.45 & 0.29 \\
     &&&&&&&&&& (0.31, 0.77) & (0.31, 0.72) & (-0.19, 0.39) & (0.25, 0.69) & (-0.12, 0.41) & (0.19, 0.67) & (-0.11, 0.64) \\ \hline
    11 &&&&&&&&&& & 0.57 & 0.42 & 0.04 & 0.32 & 0.53 & 0.21 \\
     &&&&&&&&&& & (0.34, 0.77) & (0.12, 0.67) & (-0.23, 0.31) & (-0.00, 0.59) & (0.24, 0.76) & (-0.15, 0.55) \\ \hline
    12 &&&&&&&&&&&& 0.43 & 0.73 & 0.59 & 0.63 & 0.53 \\
     &&&&&&&&&&&&  (0.21, 0.62) & (0.67, 0.91) & (0.41, 0.75) & (0.47, 0.83) & (0.25, 0.76) \\ \hline
    13 &&&&&&&&&&&&& 0.61 & NA & 0.59 & NA \\
     &&&&&&&&&&&&& (0.43, 0.80) & & (0.37, 0.80) & \\ \hline
    14 &&&&&&&&&&&&&& 0.64 & 0.67 & 0.59 \\
     &&&&&&&&&&&&&& (0.49, 0.91) & (0.53, 0.91) & (0.35, 0.91) \\ \hline
    15 &&&&&&&&&&&&&&& 0.51 & NA \\
     &&&&&&&&&&&&&&& (0.26, 0.71) & \\ \hline
    16 &&&&&&&&&&&&&&&& 0.15 \\
     &&&&&&&&&&&&&&&& (-0.20, 0.48) \\ \hline
  \end{tabular}}
 
  \label{tab:sepe}
\end{table}
\end{landscape}
\clearpage 
\section*{Figure legends}

\end{document}